\documentclass[preprint,showpacs,preprintnumbers,amsmath,amssymb]{revtex4}

\usepackage{graphicx}
\usepackage{epsfig}
\usepackage{dcolumn}
\usepackage{bm}

\newcommand{\GeV}{\mathrm{GeV}}
\newcommand{\MZ}{\mathrm{M}_Z}
\newcommand{\Zo}{\mathrm{Z}^o}
\newcommand{\cm}{\mathrm{cm}}

\begin{document}

\preprint{LNS-05-310}

\title{Mass limits for fourth generation sequential neutrinos from
  dark matter experiments}

\author{Gray Rybka}%
 \email{grybka@mit.edu}
\affiliation{%
Physics Department\\ MIT \\ Cambridge, MA 02139
}%

\author{Peter Fisher}%
 \email{fisherp@mit.edu}
\affiliation{%
Physics Department and the Laboratory for Nuclear Science\\ MIT \\ Cambridge, MA 02139
}%

\date{\today}

\begin{abstract}
Current mass limits for fourth generation sequential neutrinos come
from dark matter experiments assuming
$\rho_{DM}=0.2-0.8$
g/ cm$^3$.  We
show that the latest results from the CDMS II experiment exclude
Dirac neutrinos with masses below 500 GeV assuming only that they are
produced as expected by the Big Bang model and clump in the same
manner as baryons.  We also show the next generation of nuclear recoil experiments 
will be sensitive to fourth generation Majorana nuetrinos.  Finally, we consider the case in which the neutrino
interacts with the nucleus via the exchange of a Higgs boson.
\end{abstract}

\maketitle

\section{\label{sec:level1}Introduction}

Although not fashionable, experimental and theoretical considerations
do not exclude a fourth lepton generation.  Accelerator searches
exclude a variety charged heavy leptons up to about 100 GeV
\cite{ref:PDG2004}.  All neutrinos with masses less than $\MZ/2$ are
excluded by the $\Zo$ lineshape \cite{ref:PDG2004} and stable neutrinos
with masses less that 100 GeV are excluded assuming Dirac couplings to
$e$, $\mu$ or $\tau$.  

In the early 1990's, neutrinos with masses in the range of 10-1000 GeV
were a prime dark matter candidate \cite{ref:ahlen87,
ref:gotthard91,ref:caldwell88}.  The earliest dark
matter and solar capture measurements
excluded
neutrinos as the predominant source of dark matter assuming they had a
local density of
$\rho\sim  0.4\GeV/\cm^3$, i.e. {\em were} the dark
matter expected in our galaxy.  Since then, supersymmetric models,
most notably those with neutralinos in the 100 GeV mass range, have
taken over as the prime massive dark matter candidate
\cite{ref:ellis90,ref:jungman96}.

In the intervening time, several things have happened: 1) measurements
have shown the three lightest neutrinos are massive and
mix \cite{ref:fisher99}, 
2) advances in
cosmology have led to a much clearer understanding of the
early universe \cite{ref:freedman03}and 
3) subsequent nuclear recoil experiments have
achieved 10$^7$ greater sensitivity than the first
experiments \cite{ref:akerib04}.

In this note, we return the the question of fourth generation
sequential neutrinos as a relics of the Big Bang and use recent
experimental results to place limits on the mass of both Majorana and
Dirac neutrinos without the assumption they are the main constituent of
the galactic dark matter.  We take their calculated density
from the Big Bang and assume they fall into galaxies in the same
manner as baryons.  We assume the fourth generation neutrino, $N$, is
stable on the time scale of the universe, which means it does not mix much
with lighter species and that its partner lepton $L$ is heavier,
$m_L>m_N$.  

Our note is organized as follows: in the next section, we summarize
the production of fourth generation stable neutrinos in the Big Bang.
In the following section, we compute the recoil interaction rate and
extract limits from the current underground detectors for both
Majorana and Dirac neutrinos.  In the final section, we conclude and
discuss the limits from neutrino capture and cosmic ray
experiments.

\section{\label{sec:level2}Neutrino production in the Big Bang}

Very early in the Big Bang, when $T > 10^{10}$K, the fourth generation
neutrinos with
masses around 100 GeV would be
in equilibrium with the light charged leptons via annihilation.
However, as originally discussed in \cite{ref:weinberg77}, at about $10^{10}$ K, 
the neutrinos become sufficiently
dilute that the annihilation rate becomes small in comparison with the dilution from the
expansion of the universe.  Numerical solution to the rate
equation gives
a present day
average cosmological number density $n_o =
10^{-4}(GeV^2/cm^3)m_{\nu}^{-2}$ \cite{ref:borner03}. 

Like baryonic and dark matter, the neutrinos will concentrate in galaxies
with a typical velocity of 250 km/s.
Assuming
neutrinos clump in 
galaxies the same way baryonic matter does, this gives an estimate for
galactic neutrino 
density of $n_G=(r_G/r_D)^3 n_0$, where $r_G$ is the size of
the galaxy, and $r_D$ 
is the distance scale between galaxies.  Using 10 kpc for $r_G$ and 1
Mpc for $r_D$ 
gives $n_G=10^6 n_0$.  This estimate is good to a factor of 4, so we
can place an upper 
limit of $2 \times 10^6 n_0$ on the local heavy neutrino density from the
Big 
Bang \cite{ref:jungman96}.

\section{\label{sec:level3}Nuclear recoil detection}

\begin{figure}
\centering
\includegraphics[width=12cm]{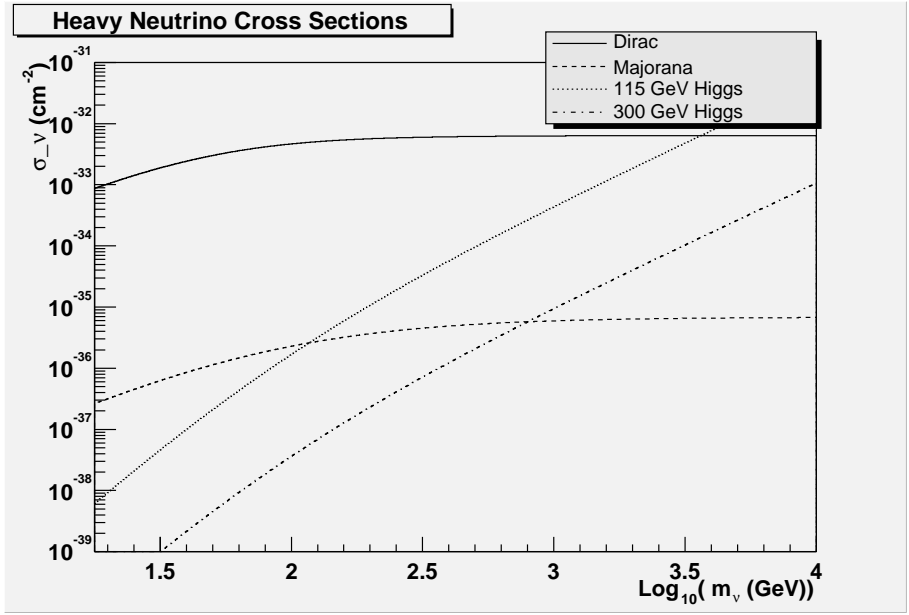}
\caption{Total Neutrino Cross Sections for Spin Dependent, Spin Independent, and Higgs Interactions}
\label{fig:totallcrossec}
\end{figure}

The spin-independent cross section for Dirac
neutrinos scattering 
coherently from a nucleus via neutral current:
\begin{eqnarray*}
\frac{d\sigma}{dT} & = &\frac{G_F^2 m_N c^2}{8 \pi v^2} 
[ Z(1-4sin^2 \theta_W)-N ]^2 \\
& \times & [1+(1-(\frac{T}{E_\nu})^2)-\frac{m_NT+m^2_\nu}{E^2_\nu}] \\
&\times & \exp(-m_N2TR^2/3\hbar^2) 
\end{eqnarray*}

where $T$ is the recoil energy, $m_N$ is the mass of the nucleus, $v$ is
the velocity of the 
neutrino, R=$1.2 \mathrm{fm} A^{\frac{1}{3}}$, and $A=N+Z$.  
The exponential term models the loss of coherence: the cross section
is down by a factor of $1/e$ when $T=3\hbar^2/2m_NR^2\sim 50
{\mathrm MeV/A}^{5/3}$.  for most target materials, this gives $T\sim 50 $ keV.
Similarly, the
cross section for the elastic scattering of
Majorana neutrinos is
\begin{equation}
\frac{d\sigma}{dT} = \frac{G_F^2 m_N c^2}{\pi v^2} C^2 \lambda^2 J(J+1)
\end{equation}
where $\lambda^2 C^2 J(J+1)$ is related to the quark spin content of
the nucleon, 
and the nucleon spin content of the nucleus \cite{ref:lewin96} Values
for 
$\lambda^2 C^2 J(J+1)$ can be found in Table \ref{table:clambda}.  
Fig.~\ref{fig:totallcrossec} shows cross sections of neutrinos of 
varying mass for different estimates. 
The neutrinos may also exchange a Higgs boson with the nucleus
\begin{equation}
    \frac{d\sigma}{dT} = \frac{G_F^2 m_N c^2}{8 \pi v^2} \frac{m_{\nu}^2 m_N^2}{m_H^4} 
    \left(1+(1-(\frac{T}{E_\nu})^2)-
\frac{m_NT+m^2_\nu}{E^2_\nu}\right)\exp\left(-\frac{m_N2TR^2}{3\hbar^2}\right)
\end{equation}
where $m_H$ is the yet unknown Higgs mass. Higgs exchange cross section will 
be the same regardless of whether the neutrino is Dirac, Majorana, or sterile.

The rate of events seen in the detector over an recoil energy range $\Delta T$ is then
\begin{equation}
R = \Delta T\frac{m_T}{m_N} \frac{\rho_\nu}{m_\nu}
\int_{v_{min}}^{v_{max}} 
\frac{d\sigma}{dT}f(v)v d^3v
\end{equation}
where $m_T$ is the target mass, $\rho_\nu$ is the local neutrino
density, 
and $f(v)$ is the Maxwell Boltzmann distribution of neutrino
velocities. We use $\overline{v}=270$ km/s for the average velocity of
the Earth relative to the dark matter halo.  
Since the recoil spectrum always falls with increasing recoil energy,
the most sensitive bin is the lowest bin above threshold. 
Thus, with an upper limit on the number of events seen in the lowest bin,
an upper 
limit on the local neutrino density can be obtained, which gives a
limit on the cosmological density $n_o$.  If the value of $n_o$
predicted for a relic neutrino of mass $m_N$ is larger than derived
from the count rate in the lowest bin in the recoil spectrum, then
neutrinos of that mass cannot exist.

The CDMS II experiment \cite{ref:akerib04} observed no counts in the  
energy range above T=20 keV with an energy resolution 
of $\Delta$T=1.5 keV over an exposure time of 19.4 kg-days. The actual
experimental
  threshold is 10 keV, but for our
  purposes the best limit is given by taking 20 keV, where the efficiency is much higher.
The 90\% confidence limit on no counts is
$N_{90}=2.3$, so the 
limit is:
\begin{equation}
\rho_\nu < \frac{N_{90}}{m_T/(m_N m_\nu) \tau \Delta T 
\int_{v_{min}}^{v_{max}} d\sigma/dT f(v)v d^3v}
\end{equation}

\begin{figure}
\centering
\includegraphics[width=9.5cm]{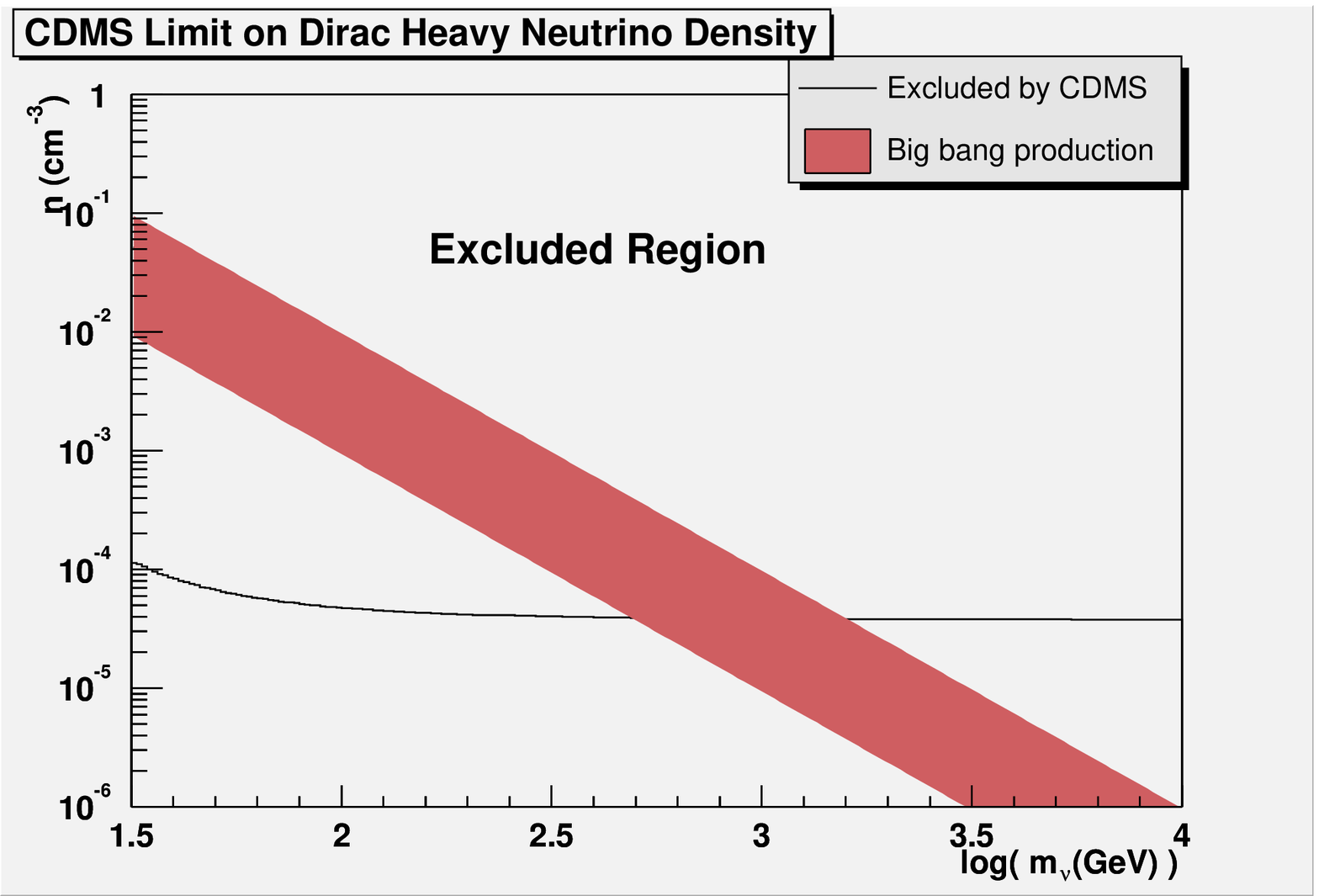}
\caption{\label{fig:diracrate}Limits on Neutrino Density for Dirac Neutrinos}
\includegraphics[width=9.5cm]{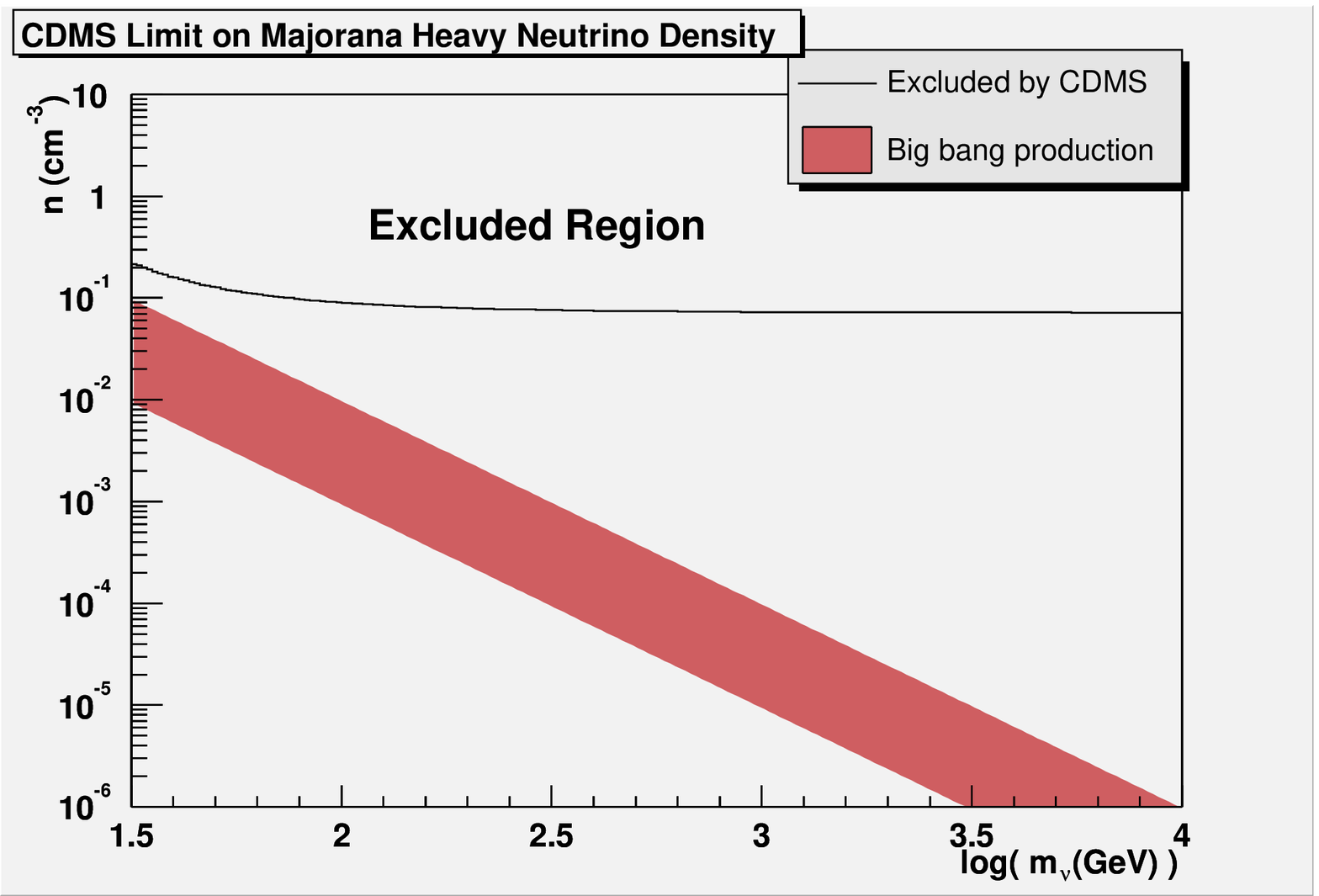}
\caption{\label{fig:rate}Limits on Neutrino Density for Majorana Neutrinos}
\includegraphics[width=9.5cm]{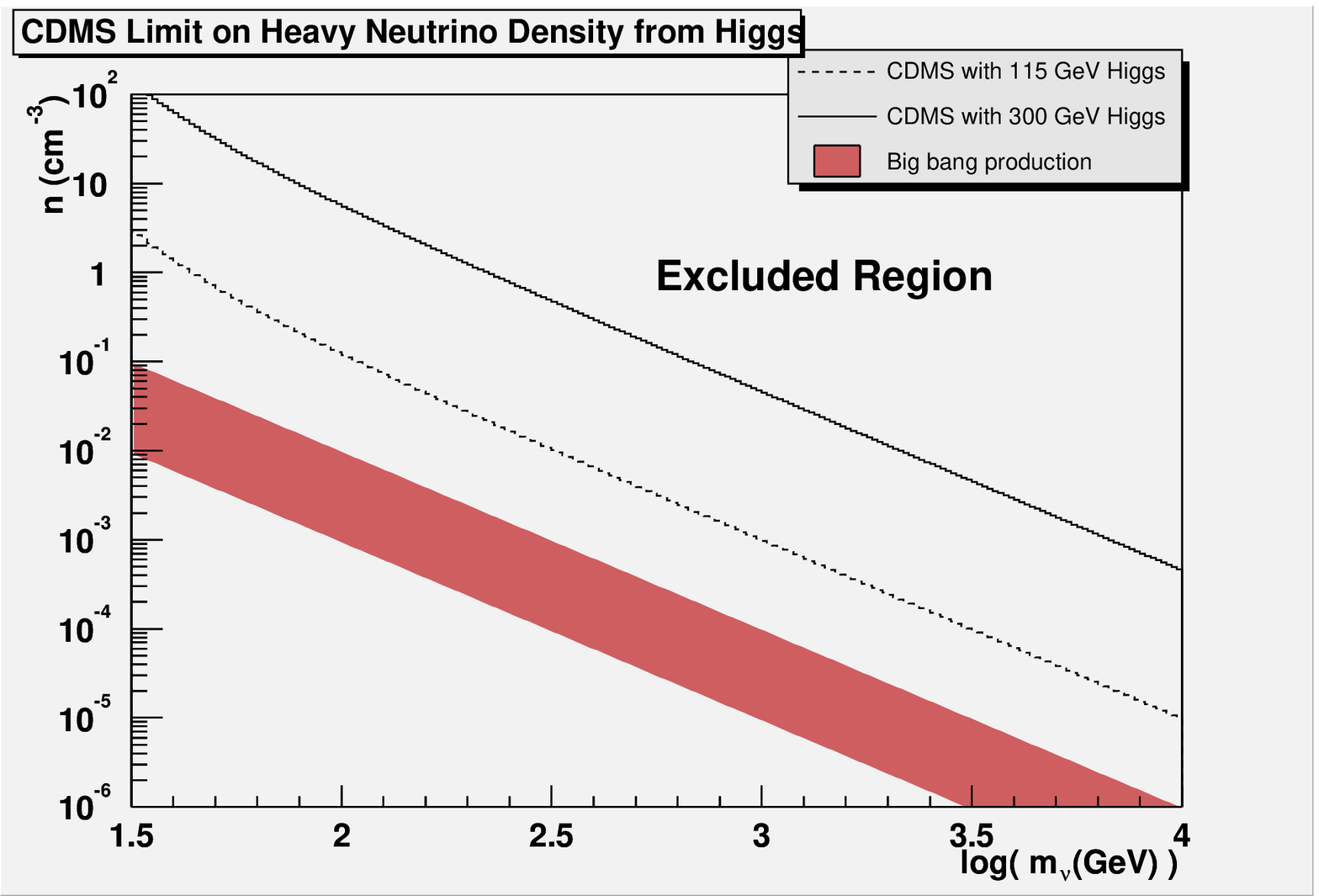}
\caption{\label{fig:higgsrate}Limits on Neutrino Density from Higgs interaction, for Higgs masses of 115GeV and 300 GeV}
\end{figure}

The 90\% confidence limits on relic neutrino density are plotted in 
Fig.~\ref{fig:rate}(a) for 
Dirac neutrinos, Fig.~\ref{fig:rate}(b) for Majorana
neutrinos, and Fig.~\ref{fig:rate}(c) for neutrinos whose interactions are dominated by Higgs exchange. 
From Fig.~\ref{fig:rate}(a), it is apparent that the CDMS II data in combination with the 
expected density of relic neutrinos from the big bang allows us to put
a lower limit 
on the mass of a possible heavy Dirac neutrino of about 500 GeV.  
If we assume a density of 0.4 g/cm$^3$, the limit increases to 3 TeV.
Fig.~\ref{fig:rate}(b) and (c) show that
current results do not exclude any of the mass range above $\MZ$/2 for Majorana neutrinos.
However, the next generation of nuclear recoil experiments will be a
factor of 10-100 more sensitive providing access to Majorana masses in
the range of 100 GeV. 
 Additionally, if the Higgs mass is found to be around 110 GeV, the dependence on
neutrino mass in the Higgs cross section would allow a detector with 
100 times the exposure of the CDMS experiment and a
null result to completely rule out the existence of heavy neutrinos of all masses.  
Conversely, if heavy neutrinos were
known to exist, a null result from such a detector could place limits on the Higgs mass.

\section{\label{sec:level5}Conclusion}

Our results assume the neutrino's density traces the density of the
baryons.  This in fact may not be true: recent numerical simulations
indicate the dark matter distribution may be rather non-uniform with
density variations as high as a factor of
thirty \cite{ref:bertschinger04}, reducing our limit to roughly 100 GeV
for Dirac neutrinos.  Since the Earth may be located in a
less dense region, there is still a loophole for massive neutrinos.
However, searches for the decay products of dark matter annihilation in cosmic rays
\cite{ref:ams04} probe the local
$\sim$3 kpc of our galaxy and solar capture 
\cite{ref:superk} signals are proportional to the average relic neutrino
density over a substantial fraction of the galaxies history.   
Both of these measurement may help close the loophole.

\begin{table*}
\caption{\label{table:clambda}
Spin content factor values obtained from Ref.~\cite{ref:lewin96}. }
\centering
\begin{tabular}{|c|c|}
\hline
Experiment & $C^2 \lambda^2 J(J+1)$ \\
\hline
NQM & 0.0260 $\pm$ 0.0013 \\
EMC 1 & 0.0221 $\pm$ 0.0020 \\
EMC 2 & 0.0169 $\pm$ 0.0046 \\
\hline
\end{tabular}
\end{table*}

\begin{acknowledgments}
GR gratefully 
acknowledges the support of the Rossi Fellowship from
the MIT Physics
Department.
\end{acknowledgments}

\newpage 
\bibliography{myrefs}

\end{document}